# Lattice Reduction Aided Precoding for Multiuser MIMO using Seysen's Algorithm

HongSun An
*Student Member, IEEE*
The Graduate School of IT & T
Inha University
Incheon, Korea
ahs3179@gmail.com

Manar Mohaisen
*Student Member, IEEE*
The Graduate School of IT & T
Inha University
Incheon, Korea
lemanar@hotmail.com

KyungHi Chang
*Senior Member, IEEE*
The Graduate School of IT & T
Inha University
Incheon, Korea
khchang@inha.ac.kr

*Abstract*—Lenstra-Lenstra-Lovasz (LLL) algorithm, which is one of the lattice reduction (LR) techniques, has been extensively used to obtain better basis of the channel matrix. In this paper, we jointly apply Seysen's lattice reduction algorithm (SA), instead of LLL, with the conventional linear precoding algorithms. Since SA obtains more orthogonal lattice basis compared to that obtained by LLL, lattice reduction aided (LRA) precoding based on SA algorithm outperforms the LRA precoding with LLL. Simulation results demonstrate that a gain of 0.5dB at target BER of $10^{-5}$ is achieved when SA is used instead of LLL for the LR stage.

## I. INTRODUCTION

Multiple-input multiple-output (MIMO) system is an attractive technology due to its capability to linearly increase the system throughput without requiring additional spectral resources [1]. The information theoretical concept of dirty paper coding has received considerable attention due to its capability to achieve the capacity region with negligible loss [2]. Therefore, when the channel state information (CSI) is available at the transmitter side by means of feedback, precoding techniques can be applied to permit to the base station to communication simultaneously with multiple users.

Zero-forcing (ZF) and minimum mean square error (MMSE) precoding are the well-known linear precoding schemes. Although linear precoding techniques have considerably low computational complexity, they have low performance due to the susceptible noise amplification, particularly when the channel matrix is ill-conditioned.

Lattice reduction (LR) is a powerful technique that can obtain better conditioned channel matrix by factorizing the channel matrix into the product of well-conditioned matrix and a unimodular matrix [3]. Thus, the motive behind applying LR algorithm is to obtain a channel matrix with shorter and more orthogonal basis. It is shown that applying lattice reduction with linear precoding schemes leads to tremendous improvement in the system performance.

Lenstra-Lenstra-Lovasz (LLL) algorithm [4] is extensively used in the literature with linear precoding schemes [5]. Also, it is shown that a better performance is achieved in case of lattice reduction-aided (LRA) precoding as compared to linear precoding schemes [6].

In this paper, we jointly apply Seysen's algorithm (SA) with linear precoding schemes. Therefore, a better conditioned channel matrix is first obtained by means of SA, over which the precoding scheme is applied. Since SA obtains more orthogonal channel compared to LLL, LRA precoding based on SA outperforms LRA precoding with LLL. In this paper, we validated this conjecture by simulation results.

The remainder of this paper is organized as follows. System model is given in Section II. In Section III, we describe Seysen's algorithm in details. In Section IV, proposed LRA precoding with linear precoding based on SA algorithm is introduced. Simulation results and discussions are given in Section V, and conclusions are drawn in Section VI.

## II. SYSTEM MODEL

We consider a downlink multi-user MIMO (MU-MIMO) system where a base station employing $N_T$ transmit antennas communicates simultaneously with $N_R$ single-antenna non-cooperative mobile stations. The system model is shown in Fig. 1.

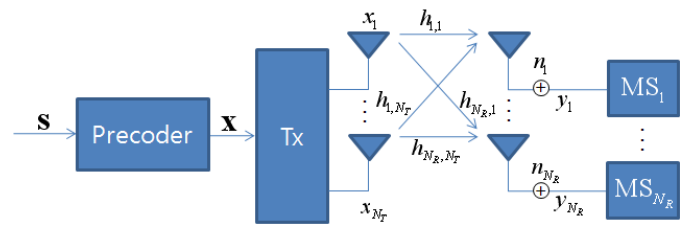

Figure 1. MU-MIMO system with precoding.

We assume a flat fading channel, where all the elements in the $N_R \times N_T$ channel matrix **H** keep constant over one frame duration. The transmitter has the full knowledge of the CSI, where the constraint on the total transmit power equals $N_T$. Let the vector **x** be the $N_T \times 1$ precoded signal, then the $N_R \times 1$ received signal vector at the $N_R$ receivers is expressed in the compact matrix form as following:

$$\mathbf{y} = \mathbf{Hx} + \mathbf{n}, \qquad (1)$$

This work was supported by the Korea Science and Engineering Foundation (KOSEF) grant funded by the Korea government (MEST) (No. R01-2008-000-20333-0).

where **n** is the $N_R \times 1$ Gaussian noise vector with elements having zero mean and variance of $\sigma_n^2$. The vector **x** is obtained by precoding the data vector **s** using a pre-filtering matrix obtained using the channel matrix.

Since LR scheme is adopted, the complex valued system model given in (1) is transformed into the equivalent real-valued system as following:

$$\mathbf{H} = \begin{bmatrix} \Re(\mathbf{H}) & -\Im(\mathbf{H}) \\ \Im(\mathbf{H}) & \Re(\mathbf{H}) \end{bmatrix}, \quad (2)$$

$$\mathbf{y} = \begin{bmatrix} \Re(\mathbf{y}) \\ \Im(\mathbf{y}) \end{bmatrix}, \quad \mathbf{x} = \begin{bmatrix} \Re(\mathbf{x}) \\ \Im(\mathbf{x}) \end{bmatrix}, \quad \mathbf{n} = \begin{bmatrix} \Re(\mathbf{n}) \\ \Im(\mathbf{n}) \end{bmatrix}, \quad (3)$$

where $\Re(\mathbf{A})$, $\Im(\mathbf{A})$ is the real part and imaginary part of **A**, respectively. And $\mathbf{H} \in \mathbb{R}^{2N_R \times 2N_T}$, $\mathbf{y}, \mathbf{n} \in \mathbb{R}^{2N_R}$, $\mathbf{x} \in \mathbb{R}^{2N_T}$.

## III. SEYSEN'S ALGORITHM

We define a lattice $L(\mathbf{H})$ whose basis vectors are $\mathbf{h}_1, \mathbf{h}_2, ..., \mathbf{h}_{2N_T}$. Then, a dual lattice $L(\mathbf{H})^*$ of the lattice $L(\mathbf{H})$, which has the relationship of $L(\mathbf{H})^* = (L(\mathbf{H})^{-1})^T$, is defined as the lattice whose basis vectors are $\mathbf{h}_1^*, \mathbf{h}_2^*, ..., \mathbf{h}_{2N_T}^*$, where $\mathbf{h}_i$ is the $i$-th column of the channel matrix **H**. Therefore, each basis vector of the lattice has the following properties:

$$\begin{aligned} (\mathbf{h}_i, \mathbf{h}_j^*) &= 1, \quad \text{for } i = j \\ (\mathbf{h}_i, \mathbf{h}_j^*) &= 0, \quad \text{otherwise} \end{aligned} \quad (4)$$

where (**a**, **b**) means inner product of two vectors. SA is proposed for simultaneously reducing the lattice $L(\mathbf{H})$ and its dual lattice $L(\mathbf{H})^*$ [7]. We define **A** and **A**\* as the associated quadratic forms of $L(\mathbf{H})$ and $L(\mathbf{H})^*$, respectively. Since $\mathbf{A} = \mathbf{H}^T\mathbf{H}$, the element $a_{ij}$ of matrix **A** is the inner product of the basis vector $\mathbf{h}_i$ and $\mathbf{h}_j$ of the lattice $L(\mathbf{H})$

$$\begin{aligned} \mathbf{A} &= [a_{ij}] = [(\mathbf{h}_i, \mathbf{h}_j)], \\ \mathbf{A}^* &= [a_{ij}^*] = [(\mathbf{h}_i^*, \mathbf{h}_j^*)], \quad \text{for } 1 \leq i, j \leq 2N_T. \end{aligned} \quad (5)$$

Let the lattice $L(\mathbf{H})$ be generated by **H**, i.e., the columns of **H** are the basis vectors of $L(\mathbf{H})$, then any other generating matrix $\tilde{\mathbf{H}}$ of the lattice $L(\mathbf{H})$ can be written as

$$\tilde{\mathbf{H}} = \mathbf{HT}, \quad (6)$$

where **T** is an $2N_T \times 2N_T$ integer unimodular matrix, with a unity determinant, i.e., det(**T**) = ±1 [8]. From (5), the quadratic form of $\tilde{\mathbf{H}}$ can be written as:

$$\tilde{\mathbf{A}} = \mathbf{T}^T \mathbf{A} \mathbf{T}. \quad (7)$$

The Seysen's measure of the quadratic form **A** can be defined as following:

$$S(\mathbf{A}) = \sum_{i=1}^{2N_t} a_{ii} a_{ii}^* = \sum_{i=1}^{2N_t} \|\mathbf{h}_i\|^2 \|\mathbf{h}_i^*\|^2. \quad (8)$$

The Seysen's measure has minimum value of $2N_T$ iff the basis vectors of the obtained matrix are orthogonal. Therefore, SA tries iteratively to attain this bound by applying arithmetic operations on the columns, i.e., lattice basis vectors, of the channel matrix. A reduced lattice of $L(\mathbf{H})$ and its corresponding quadratic form **A** are called S-reduced if Seysen's measure $S(\mathbf{A})$ cannot be reduced anymore. This is expressed as following:

$$S(\mathbf{A}) \leq S(\mathbf{T}^T \mathbf{A} \mathbf{T}), \quad \text{for all } \mathbf{T}. \quad (9)$$

To obtain the optimal transformation matrix **T** for a given lattice $L(\mathbf{H})$, the class of transformation matrices **T** is defined by

$$\begin{aligned} \mathbf{T}_{i,j}^{\lambda_{ij}} &= \mathbf{I}_{2N_T} + \lambda_{ij} \mathbf{U}_{ij} \quad i \neq j, \; \lambda \in \mathbb{Z} \\ \mathbf{U}_{ij} &= [u] \quad u = \begin{cases} 1, & \text{only } (i,j) \text{ pair,} \\ 0, & \text{otherwise,} \end{cases} \end{aligned} \quad (10)$$

where the matrix $\mathbf{I}_{2N_T}$ is the $2N_T$ identity matrix, the matrix $\mathbf{U}_{ij}$ has only one nonzero element, $\mathbf{T}_{ij}^{\lambda_{ij}}$ is a matrix with all the diagonal elements are 1s, and only one nonzero off-diagonal element, and $\lambda_{ij}$ has the following value:

$$\lambda_{ij} = round\left\{\frac{1}{2} \cdot \left(\frac{a_{i,j}^*}{a_{j,j}^*} - \frac{a_{i,j}}{a_{i,i}}\right)\right\}, \quad (11)$$

where $\lambda_{ij}$ minimizes Seysen's measure for the chosen $(i, j)$ column pair [8].

By multiplying the matrix $\mathbf{T}_{ij}^{\lambda_{ij}}$, defined in (10), to the right side of the **H**, we can simply get the transformed basis vectors $\mathbf{h}_j' = \mathbf{h}_j + \lambda_{ij}\mathbf{h}_i$. Because it is easy to calculate the transformation matrix $\mathbf{T}_{ij}^{\lambda_{ij}}$, the final transformation matrix **T** is calculated directly from the obtained $\mathbf{T}_{ij}^{\lambda_{ij}}$ matrices as following:

$$\mathbf{T} = \prod_k \mathbf{T}_{ij}^{\lambda_{ij}}, \quad \text{for } 1 \leq k < \infty, \quad (12)$$

where $k$ is the number of bases transformation till the Seysen' measure can't be reduced anymore. In this case, the quadratic form **A** is called $S_2$-reduced and then presented as

$$S(\mathbf{A}) \leq S\left((\mathbf{T}_{ij}^{\lambda_{ij}})^T \mathbf{A} \mathbf{T}_{ij}^{\lambda_{ij}}\right), \quad (13)$$
$$\text{for } \lambda_{ij} \in \mathbb{Z} \text{ with } 1 \leq i, j \leq 2N_T.$$

The choice of the columns pair $(i, j)$ at any iteration plays a very important role in the convergence and the complexity of the SA. Therefore, a better choice of this pair can lead to reduced number of iterations while obtaining the same performance of a random choice. Based on this discussion, we have two SAs which differ in the way the pair $(i, j)$ is selected, namely; lazy and greedy SAs.

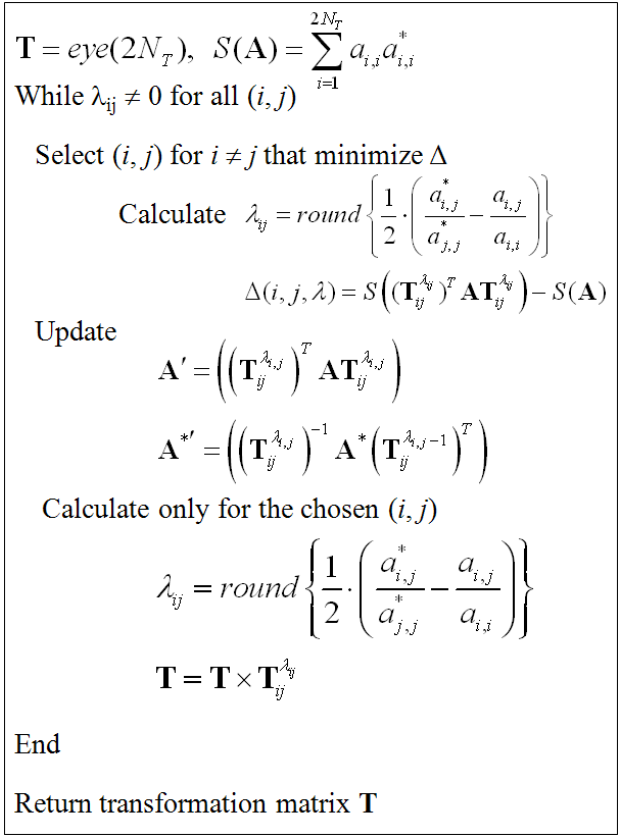

Figure 2. Seysen's greedy algorithm.

In lazy SA method, column pair is selected randomly from all the pairs $(i, j)$, such that $\lambda_{ij} \neq 0$. This choice is repeated iteratively until the quadratic form **A** is S-reduced. On the other hand, greedy selection method selects a pair $(i, j)$ that minimizes $\Delta$, where $\Delta$ is given by:

$$\Delta(i, j, \lambda) = S\left((\mathbf{T}_{ij}^{\lambda_{ij}})^T \mathbf{A} \mathbf{T}_{ij}^{\lambda_{ij}}\right) - S(\mathbf{A}). \quad (14)$$

The largest reduction of $\Delta$ leads to the largest decrease of Seysen's measure. One disadvantage of the Greedy selection is that it requires additional computations to get $\Delta$. However, it can reduce the number of iterations [8], [9].

The algorithmic description of SA greedy LR method is given in Fig. 2.

## IV. LATTICE REDUCTION AIDED PRECODING

The basic concept of the proposed LRA precoding is described in Fig. 3. In order to jointly apply LR technique with precoding schemes, we apply LR to $\mathbf{H}^T$ as following

$$\tilde{\mathbf{H}} = \mathbf{T}^T \mathbf{H}, \quad (15)$$

where $\tilde{\mathbf{H}}$ is row-wise more orthogonal than **H**, and **T** is a unimodular matrix which has the properties mentioned in (4). Linear precoding techniques lead to noise amplification which is seen as increasing the transmit power requirements. Therefore, to prevent this noise amplification, LRA precoding uses $\tilde{\mathbf{H}}$ instead of **H** leading to reduced amplification of the noise. In this paper, we combine ZF (Channel Inversion) and MMSE (Regularized Channel Inversion) precoding schemes with Seysen's lattice reduction technique.

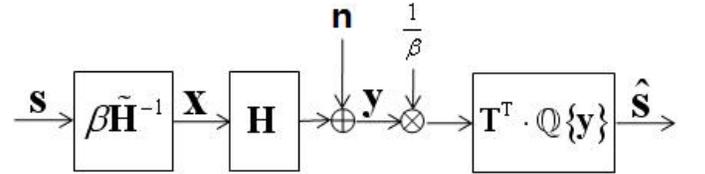

Figure 3. LRA precoding.

### A. Lattice Reduction Aided Linear Precoding using SA

*1) Linear ZF Precoding with SA LR*

Linear ZF precoding is a well-known linear precoding technique that pre-filters the transmitted symbols using the pseudo-inverse of the channel matrix. The precoded vector is defined as $\mathbf{x} = \mathbf{W}\mathbf{s}$, where **W** is defined as following

$$\mathbf{W} = \beta \mathbf{H}^{-1},$$
$$\text{where } \beta = \sqrt{\frac{N_T}{trace(\mathbf{H}^T \cdot \mathbf{H})^{-1}}}. \quad (16)$$

In (16), the scaling factor $\beta$ is used to restrict the total transmit power to the predefined limit $N_T$. At the receiver side, the symbols are recovered by dividing the received vector by $\beta$. In case of MU-MIMO systems, inter-user interference can be therefore totally canceled by using this approach. Nonetheless, if the channel is ill-conditioned, serious noise amplification arises. The application of LR before the linear ZF precoding leads to tremendous reduction in $\beta$, which consequently reduces the noise amplification. Since that SA makes more orthogonal lattice basis compared to LLL algorithm, a better performance can be obtained.

*2) Linear MMSE Precoding with SA LR*

Linear MMSE precoding schemes regularizes the channel matrix to reduce the transmit power. This regularization leads to inter-user interference, while reduces the noise amplification. To improve the BER performance, we apply SA LR with the conventional linear MMSE precoding algorithm. At first, the channel matrix is extended as following:

$$\mathbf{H}_{ex} = [\mathbf{H}_{real} \ \sigma_n \mathbf{I}_M], \quad (17)$$

where $\sigma_n$ is the noise standard deviation. SA algorithm is applied to $\mathbf{H}_{ex}^T$ as in (18).

$$\tilde{\mathbf{H}} = \mathbf{T}^T \mathbf{H}_{ex} \quad (18)$$

From (18), the linear MMSE precoding is applied using $\tilde{\mathbf{H}}$. The procedure is described as following:

$$\left.\begin{array}{l} \mathbf{x}' = \beta \tilde{\mathbf{H}}^\dagger \mathbf{s}, \ \mathbf{x} = \mathbf{x}'(1:N/2) \quad \text{at Tx side} \\ \mathbf{y} = \mathbf{H}\mathbf{x} + \mathbf{n} = \beta \mathbf{H}\mathbf{H}_{ex}^\dagger \mathbf{T}^{-T}\mathbf{s} \\ \hat{\mathbf{s}} = \mathbf{T}^T \mathbb{Q}\{\mathbf{y}\} \end{array}\right\} \quad \text{at Rx side}, \quad (19)$$

where $(\cdot)^\dagger$ is the pseudo-inverse, $N$ is equal to $2 \times N_T$, and $\mathbf{x}$ is the first $N/2$ elements of $\mathbf{x}'$. Also, $\beta$ is calculated using (16) where $\mathbf{H}$ is replaced by $\tilde{\mathbf{H}}$. The combination of the introduced SA LR with the linear MMSE precoding leads to tremendous reduction in the transmit power which reduces the noise amplification and consequently improves the performance.

## V. SIMULATION RESULTS

In this Section, we evaluate the bit error rate performance of the lattice reduction aided linear precoding. We use both linear ZF and MMSE precoding schemes with the conventional LLL and the introduced SA. We consider a MU-MIMO system, where the BS has 4 transmit antennas and simultaneously communicates with 4 single-antenna non-cooperating users. Transmitted symbols are drawn from a 4QAM constellation set. The BS has perfect knowledge of CSI by means of feedback from the users.

Fig. 4 depicts the cumulative distribution function of the condition number of the precoding matrix in case of LLL, SA, and without LR. These results are obtained by averaging the condition number of the resulting matrices over 10,000 independent realization of the channel matrix. The condition number is defined as following:

$$\kappa(\mathbf{H}) = \frac{\sigma_{\max}(\mathbf{H})}{\sigma_{\min}(\mathbf{H})} \geq 1, \quad (20)$$

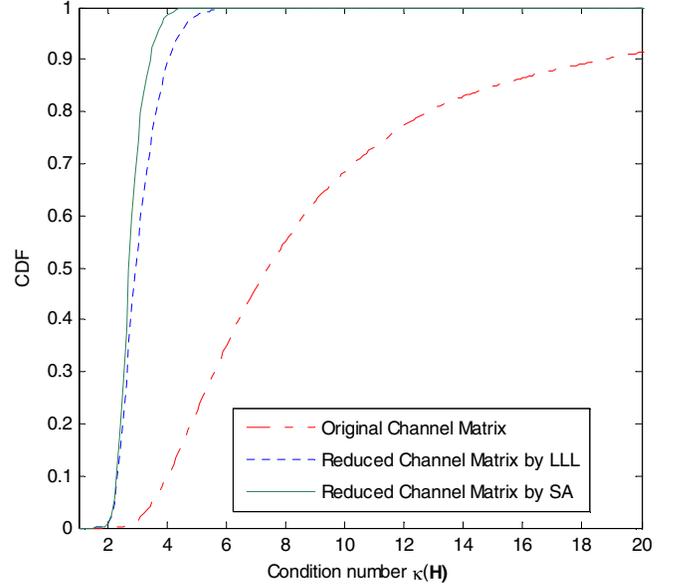

Figure 4. CDF of condition number for the reduced matrix by SA and LLL schemes.

where $\sigma_{\max}(\mathbf{H})$ and $\sigma_{\min}(\mathbf{H})$ are the maximal and the minimal singular value of $\mathbf{H}$, respectively. Hence, if the condition number of the channel matrix is close to 1, channel matrix becomes more orthogonal leading to reduction in the noise amplification. On the other hand, if the condition number is large, the channel becomes ill-conditioned and the noise amplification problem clearly arises. From Fig. 4, one can conclude that SA LR leads to better conditioned channel matrix compared to that obtained by the LLL LR algorithm and much better than that of the original channel matrix without lattice reduction.

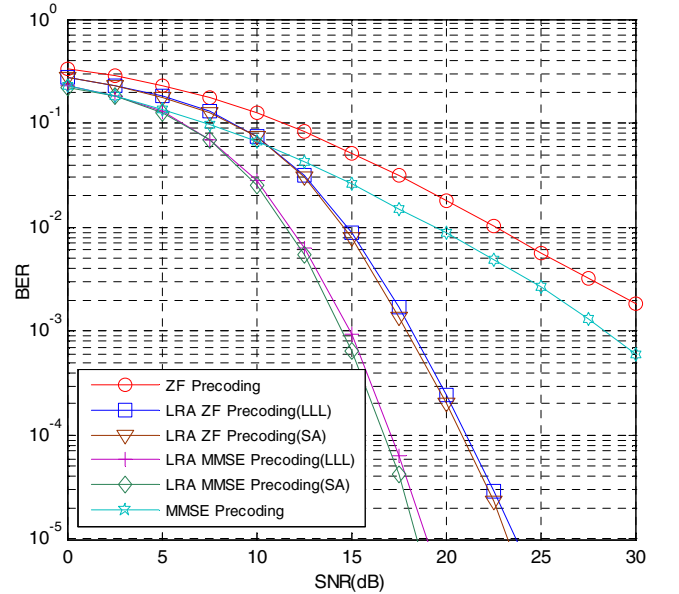

Figure 5. BER performances of the LRA linear precoding schemes.

Fig. 5 depicts the uncoded BER performance of SA and LLL based LRA precoding for same system configuration using 4QAM modulation. The elements of **H** are modeled as independent and identically distributed circularly-symmetric complex Gaussian random variables. Linear precoding jointly applied with LLL LR algorithm clearly outperforms the linear precoding. At a target BER of $10^{-2}$, the gain in the transmission power is 7.5dB. Also, our introduced joint precoding of SA LR and linear precoding outperforms the conventional LRA precoding using LLL for the whole range of SNR values. For instance, at a target BER of $10^{-5}$, SA LRA precoding outperforms LLL LRA precoding by 0.5dB, while SA has computational complexity approximately 92% of LLL algorithm for 4×4 MIMO system using 4QAM [9], [10].

## VI. Conclusions

In this paper, we introduced an LRA precoding technique based on Seysen's lattice reduction algorithm for MU-MIMO systems. LLL and the introduced SA LR techniques are jointly applied with linear precoding schemes. We showed that the application of the LR techniques lead to tremendous gains in the performance, where a gain of about 7.5dB and 8dB are obtained in the case of ZF and MMSE precoding, respectively. Moreover, SA is shown to obtain better conditioned channel matrix compared to LLL, which improves the BER performance. As a result, LRA linear precoding with SA outperforms LLL based LRA linear precoding. In light of the above, SA is shown to be a prominent candidate as a lattice reduction scheme for MU-MIMO systems with precoding.